\documentclass{article}
\usepackage{spconf,amsmath,graphicx, enumitem}
\usepackage[colorlinks=true,      
linkcolor=blue,       
urlcolor=blue,        
citecolor=blue]{hyperref} 
\usepackage{amsthm,amssymb}
\usepackage{mathrsfs}
\usepackage{booktabs} 
\usepackage{multirow} 
\usepackage{caption}    
\usepackage{pgfplots}
\usepgfplotslibrary{groupplots}
\pgfplotsset{compat=1.18}

\title{Whisper-MLA: Reducing GPU Memory Consumption of ASR Models based on MHA2MLA Conversion}
\name{\textnormal{\begin{tabular}{c} Sen Zhang$^{1}$, Jianguo Wei$^{1}$, Wenhuan Lu$^{1}$, Xianghu Yue$^{1, \dagger}$, \\ Wei Li$^{2}$, Qiang Li$^{2}$, Pengcheng Zhao$^{2}$, Ming Cai$^{2}$, Luo Si$^{2}$ \end{tabular}} \thanks{$^{\dagger}$ Corresponding author.}
\thanks{Our source code is publicly available at \href{https://github.com/ssss-sen/Whisper_MLA}{https://github.com/ssss-sen/Whisper\_MLA}.}}
\address{$^{1}$College of Intelligence and Computing, Tianjin University, Tianjin, China\\
$^{2}$Banma Network Technology Co., Ltd.}
\begin{document}
%
\maketitle
\begin{abstract}
The Transformer-based Whisper model has achieved state-of-the-art performance in Automatic Speech Recognition (ASR). However, its Multi-Head Attention (MHA) mechanism results in significant GPU memory consumption due to the linearly growing Key-Value (KV) cache usage, which is problematic for many applications especially with long-form audio. To address this, we introduce Whisper-MLA, a novel architecture that incorporates Multi-Head Latent Attention (MLA) into the Whisper model. Specifically, we adapt MLA for Whisper's absolute positional embeddings and systematically investigate its application across encoder self-attention, decoder self-attention, and cross-attention modules. Empirical results indicate that applying MLA exclusively to decoder self-attention yields the desired balance between performance and memory efficiency. Our proposed approach allows conversion of a pretrained Whisper model to Whisper-MLA with minimal fine-tuning. Extensive experiments on the LibriSpeech benchmark validate the effectiveness of this conversion, demonstrating that Whisper-MLA reduces the KV cache size by up to 87.5\% while maintaining competitive accuracy.
\end{abstract}
\section{Introduction}
\label{sec:intro}

The Transformer architecture, anchored by its Multi-Head Attention (MHA) mechanism, has driven substantial advances in Automatic Speech Recognition (ASR) systems \cite{gulati2020conformer, park2019specaugment, dong2018speech}. OpenAI’s Whisper model \cite{radford2023robust}, for instance, has achieved state-of-the-art accuracy across diverse acoustic conditions. However, this powerful architecture incurs significant GPU memory consumption, a limitation that becomes particularly pronounced when processing long-form audio input \cite{10.1145/3530811}. 

A key bottleneck for long-context tasks is the Key-Value (KV) cache inherent in the MHA mechanism \cite{vaswani2017attention}, which scales linearly with both sequence length and model size \cite{kwon2023efficient}. To address this, DeepSeek \cite{liu2024deepseek} introduced Multi-Head Latent Attention (MLA), an attention mechanism that employs low-rank key-value joint compression. MLA significantly reduces the KV cache during inference and has been shown to outperform standard MHA on various text-based benchmarks.

To facilitate the adoption of MLA, the MHA2MLA framework \cite{ji2025towards} was proposed to convert mainstream MHA-based Large Language Models (LLMs) to the MLA architecture without requiring full pretraining from scratch. By leveraging techniques like partial rotary position embedding (partial RoPE) and low-rank approximation, MHA2MLA maximizes parameter reuse from pretrained models, thereby reducing inference costs while preserving performance.

\begin{figure*}[ht]
	\centering 
	\includegraphics[width=1\textwidth]{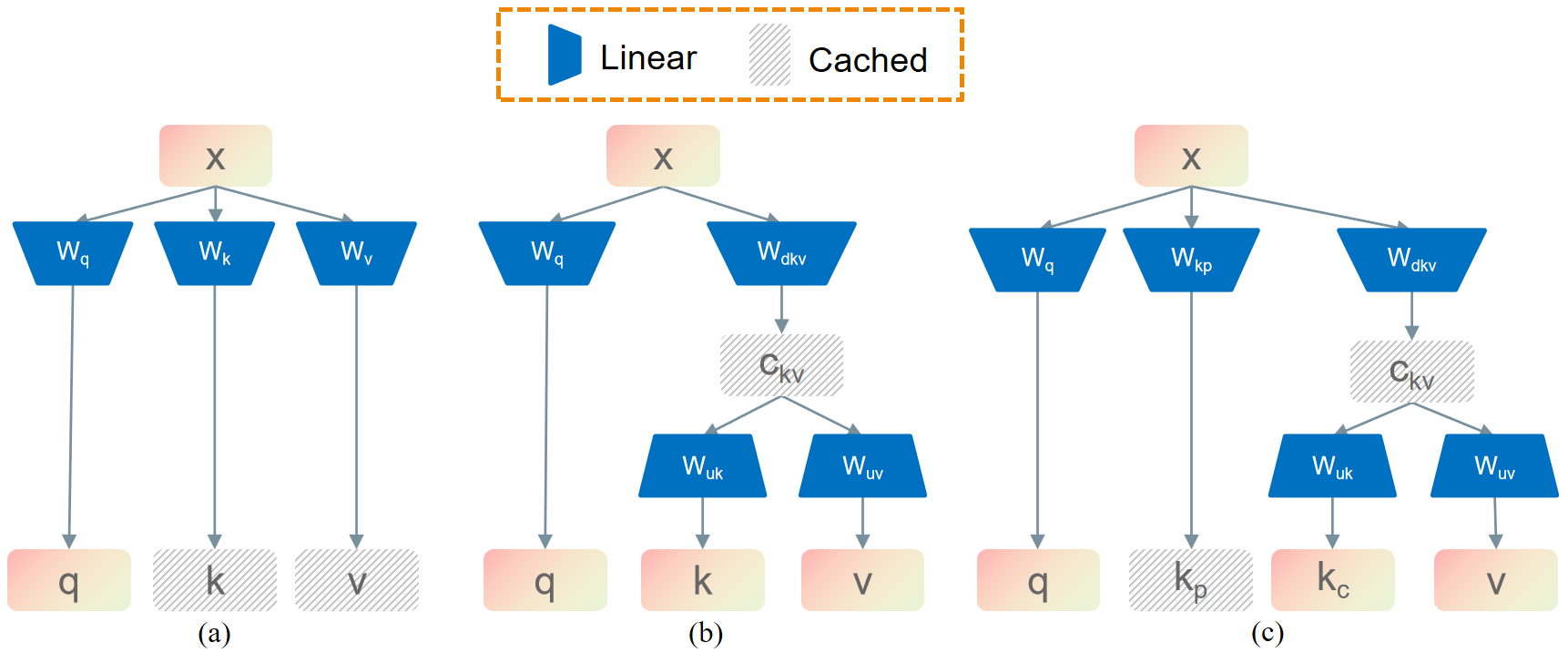} 
	\caption{Three attention architectures: (a) Original MHA in Whisper (left), (b) Full low-rank compression MLA (middle), (c) Dimension-preserving MLA (right).} 
	\label{Figure mla} 
\end{figure*}

In the domain of speech processing, the memory challenge is further exacerbated, making efficient attention mechanisms particularly critical \cite{zhang2020transformer}. To the best of our knowledge, while MLA has proven effective for textual modalities, its application to encoder-decoder ASR architectures remains largely unexplored. This motivates our work to adapt MLA for ASR, specifically targeting the Whisper model to reduce its inference memory footprint.

However, the MHA2MLA framework is incompatible with Whisper due to fundamental architectural differences. The standard MLA formulation is predicated on relative positional encodings \cite{touvron2023llama, bai2023qwen}, such as Rotary Positional Encoding (RoPE) \cite{su2024roformer}, where keys and queries are decoupled into "content" and "position" vectors to explicitly inject positional information. Directly applying this architecture to Whisper, which employs absolute positional encodings with pre-integrated positional features, would break the consistency of the attention mechanism.

To address this challenge, we propose a novel MLA design tailored for Whisper.
Our approach preserves the model's original parameters and capabilities to the greatest extent by retaining its absolute positional encoding schemes. Moreover, while MLA has been applied only to self-attention in decoder-only models, we explore its effectiveness within encoder-decoder framework. In particular, we systematically investigate the integration of MLA into encoder self-attention, decoder self-attention, and decoder cross-attention modules, respectively.

Our main contributions are summarized as follows:
\begin{itemize}
    \item We propose a novel MLA architecture for models with absolute positional encoding, enabling substantial reductions in memory consumption during inference while preserving recognition performance.
    \item We conduct a systematic study of MLA deployment across different attention modules in Whisper and present Whisper-MLA, a variant that achieves an effective balance between memory efficiency and accuracy.
    \item We develop a parameter-efficient conversion strategy that adapts a pre-trained Whisper model to Whisper-MLA with minimal fine-tuning, eliminating the need for costly training from scratch.
\end{itemize}



\section{Methodology}
\label{sec:method}

\subsection{MLA structure adapted for Whisper}
\label{ssec:mla structure}
 
Since query projections do not contribute to the KV cache size during autoregressive inference, we retain Whisper's original query processing scheme. For keys and values, we introduce two MLA-based processing approaches, as depicted in \autoref{Figure mla}. Analysis in \autoref{sec:result} reveals that while compressing all key and value dimensions into a low-rank latent space is possible (Figure 2b), preserving a small subset of original key dimensions from compression significantly improves ASR performance (Figure 2c). This dimension-preserving design also aligns more closely with the original MLA concept.

The Whisper encoder employs sinusoidal positional embeddings, where consecutive dimension pairs form a frequency subspace, encoding position with a sine and cosine of a specific frequency \cite{vaswani2017attention}. Therefore, our dimension preservation strategies operate on these subspaces rather than individual dimensions. Given a requirement to retain $r$ frequency subspaces, we employ two selection strategies:

{\bf Uniform sampling}: This approach balances high- and low-frequency components by selecting $r$ subspaces with the same geometrically spaced intervals:
\begin{equation}
	D_{\text{uniform}} = \{ \lfloor k\frac{d_h}{2r} \rfloor \mid 0 \leq k < r \}
	\tag{1}
\end{equation}

{\bf Head-wise 2-norm contribution}: This strategy ranks subspaces based on the insight that the product of the 2-norms of the query and the key vectors serves as an upper bound on their contribution to the attention score \cite{barbero2410round}. Following this insight, we quantify the utility of each frequency subspace by computing the average 2-norm product over multiple training samples. The $r$ subspaces with the highest scores are selected:
\begin{equation}
    	D_{\text{2-norm}} = \text{top-r}_{0 \leq k < \frac{d}{2}}(\Arrowvert \mathbf{q}_*^{[2k, 2k+1]} \Arrowvert \Arrowvert \mathbf{k}_*^{[2k, 2k+1]} \Arrowvert)
    	\tag{2}
    \end{equation}

We compare full compression and both dimension-preserving strategies in \autoref{sec:result}.

\subsection{The overall architecture of Whisper-MLA}
\label{ssec:overall architecture}

We propose and evaluate two Whisper-MLA architectures:

{\bf Whisper-MLA (Full)}: We convert all attention mechanisms in the Whisper model (including encoder self-attention, decoder self-attention, and cross-attention) to the MLA structure. This configuration tests the general applicability of MLA across different types of attention.

{\bf Whisper-MLA (Decoder Self-attention Only, DSO)}:
This configuration applies MLA exclusively to the decoder's self-attention, a targeted approach designed to balance memory efficiency with ASR performance. First, the primary bottleneck during inference is the dynamically growing KV cache, an issue unique to the decoder's self-attention. The encoder's KV cache, in contrast, is static after a single forward pass. Consequently, modifying only the decoder self-attention achieves the same KV cache reduction as a full model conversion. Second, this configuration preserves ASR performance by leaving the Whisper encoder unaltered. Given its pre-training on a massive dataset, the encoder's parameters are highly optimized for extracting robust acoustic representations and are widely adopted in various models \cite{tang2023salmonn, xie2024mini, fang2024llama}. Altering its self-attention mechanism would risk degrading this critical capability. By leaving the encoder and cross-attention untouched, our DSO approach mitigates the memory issue without disrupting critical acoustic modeling.

\begin{figure}
	\centering 
	\includegraphics[width=0.49\textwidth]{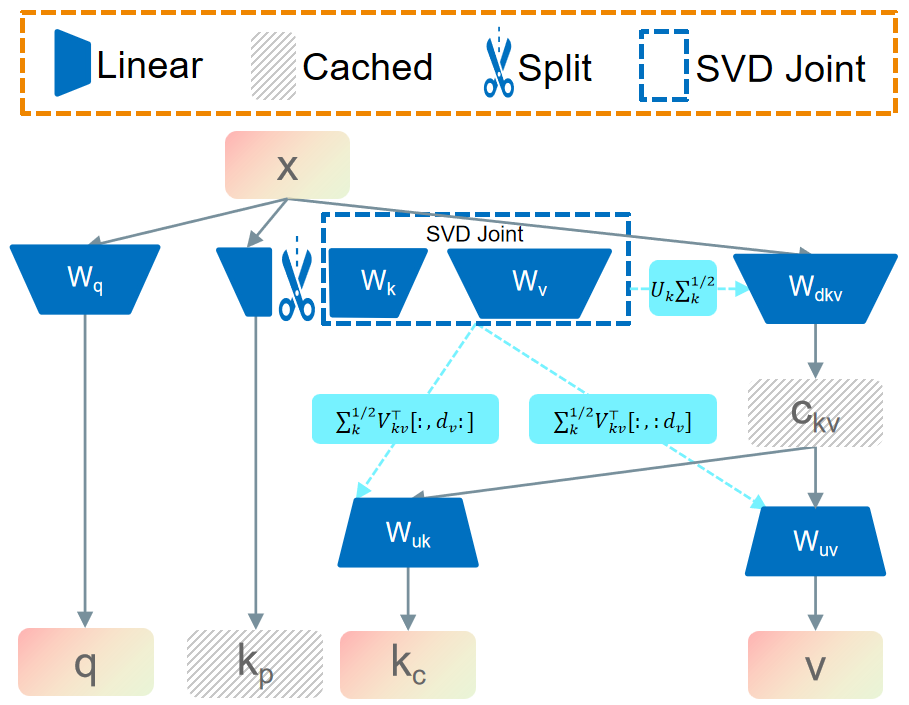} 
	\caption{The method of converting Whisper to Whisper-MLA.} 
	\label{Figure conversion} 
\end{figure}

\subsection{Conversion from Whisper to Whisper-MLA}
\label{ssec:conversion}
Following the methodology of MHA2MLA \cite{ji2025towards}, we propose a parameter-efficient fine-tuning framework that adapts pretrained Whisper to Whisper-MLA without training from scratch. As illustrated in \autoref{Figure conversion}, we leave the query projection $\mathbf{W}_q$ unchanged. We split the key projection $\mathbf{W}_k$ into a preserved part $\mathbf{W}_{kp}$ and a compressible part $\mathbf{W}_{kc}$ based on the dimension selection strategies from \autoref{ssec:mla structure}. We then perform a joint Singular Value Decomposition (SVD) on the concatenated matrix $[\mathbf{W}_{kc}, \mathbf{W}_v]$ to obtain a low-rank approximation.

The joint SVD factorizes the concatenated matrix as:
\begin{equation}
	[\mathbf{W}_{kc}, \mathbf{W}_v] = \mathbf{U}_{kv} \mathbf{\Sigma}_{kv} \mathbf{V}_{kv}^\top 
	\tag{3}
\end{equation}
where $\mathbf{U}_{kv}, \mathbf{V}_{kv} \in \mathbb{R}^{d_h \times d_{kv}}$, $\mathbf{\Sigma}_{kv} \in \mathbb{R}^{d_{kv} \times d_{kv}}$, the latent projection is then defined as:
\begin{equation}
	\mathbf{W}_{uk} = \mathbf{\Sigma}_{kv}^{1/2}\mathbf{V}_{kv}[:, d_v:], \mathbf{W}_{uv} = \mathbf{\Sigma}_{kv}^{1/2}\mathbf{V}_{kv}[:, :d_v]
	\tag{4}
\end{equation}

This approach jointly optimizes a shared latent space for the compressible key components and the values, maximizing parameter reuse from the pretrained Whisper model.

\section{Experimental setup}
\label{sec:experiment}

We employ the official Whisper-small model as our baseline, which contains 244 million parameters and was pre-trained on 680,000 hours of speech data \cite{radford2023robust}. Leveraging DeepSeek's empirical heuristics \cite{liu2024deepseek}, our dimension-preserving strategy maintains 48 of the 768 key dimensions and applies low-rank approximation to project the remaining key and all value dimensions into a unified latent space of size 96.

We convert the pretrained Whisper model to Whisper-MLA and then fine-tune it on the 960-hour LibriSpeech dataset \cite{panayotov2015librispeech} for 3 epochs. 
We evaluate our model on the LibriSpeech validation and test sets.

All experiments were conducted on a single NVIDIA RTX 4090 GPU (24GB) with a batch size of 8 and a gradient accumulation factor of 4. Under this configuration, converting Whisper to Whisper-MLA requires only 12 hours, highlighting the practical advantage of our method in resource-constrained scenarios.

\begin{table*}[htbp] 
	\centering 
	\caption{WER (\%) on LibriSpeech and KV Cache Memory Reduction for Whisper and Whisper-MLA Models.}
	\label{tab:wer_comparison}
	\renewcommand{\arraystretch}{1.0}
	\begin{tabular*}{\textwidth}{@{} c @{\extracolsep{\fill}} ccccccc @{}}
		\toprule
		\textbf{Model} & \textbf{Keep Strategy} & \textbf{KV Cache Red.} & \textbf{dev-clean} & \textbf{dev-other} & \textbf{test-clean} & \textbf{test-other} & \textbf{avg} \\
		\midrule
		
		Whisper(pretrained) & - & 0\% & 16.37 & 19.78 & 16.00 & 20.90 & 18.36 \\
		\midrule
		Whisper(finetuned) & - & 0\% & \textbf{6.32} & \textbf{14.86} & 6.86 & \textbf{15.05} & \textbf{10.95} \\
		\midrule
		
		\multirow{3}{*}{\begin{tabular}[c]{@{}c@{}}Whisper-MLA (Full)\end{tabular}} 
		& Full Compression & \textbf{87.50\%} & 16.58 & 27.79 & 16.37 & 28.09 & 22.46 \\
		& Uniform       & 81.25\% & 12.32 & 19.72 & 13.18 & 21.34 & 16.81 \\
		& 2-Norm        & 81.25\% & 11.75 & 20.83 & 12.15 & 21.74 & 16.82 \\
		\midrule
		
		\multirow{3}{*}{\begin{tabular}[c]{@{}c@{}}Whisper-MLA (DSO)\end{tabular}} 
		& Full Compression & \textbf{87.50\%} & 8.69 & 16.99 & 8.87 & 17.86 & 13.29 \\
		& Uniform       & 81.25\% & 6.60 & 15.23 & \textbf{6.61} & 15.32 & 11.12 \\
		& 2-Norm        & 81.25\% & 7.33 & 16.17 & 7.82 & 16.18 & 12.06\\
		
		\specialrule{1.2pt}{0pt}{0pt}
	\end{tabular*}
\end{table*}

\begin{figure*}[ht] 
    \centering
    \begin{tikzpicture}
        \begin{groupplot}[
            group style={
                group size= 4 by 1,
                horizontal sep=0.75cm, 
                vertical sep=1.5cm,   
                xlabels at=edge bottom,
                ylabels at=edge left
            },
            width=0.29\linewidth,     
            height=0.3\linewidth,    
            ybar=0.1pt,
            enlargelimits=0.15,
            legend style={at={(2.3,1.25)}, anchor=north, legend columns=-1, font=\small,
            },
            ylabel={GPU Memory Consumption(GB)},
            xlabel={sequence length},
            symbolic x coords={256, 512, 1024, 2048, 4096},
            xtick=data,
            tick label style={font=\small}, 
            label style={font=\small}       
        ]
        
        \nextgroupplot[bar width=8pt]
        \addplot+[area legend] coordinates {(256, 1.15) (512, 1.15) (1024, 1.15) (2048, 1.31)(4096, 1.59)};
        \addplot+[area legend] coordinates {(256, 1.01) (512, 1.01) (1024, 1.01) (2048, 1.02)(4096, 1.19)};
        \legend{Whisper, Whisper-MLA}
        \node[anchor=north west] at (rel axis cs:0.02,1) {\textbf{bsz=1}};
        
        \nextgroupplot[bar width=8pt]
        \addplot coordinates {(256, 1.83) (512, 1.83) (1024, 1.83) (2048, 2.49) (4096, 3.62)};
        \addplot coordinates {(256, 1.7) (512, 1.7) (1024, 1.7) (2048, 1.7) (4096, 2.2)};
        \node[anchor=north west] at (rel axis cs:0.02,1) {\textbf{bsz=4}};
        
        \nextgroupplot[bar width=8pt]
        \addplot coordinates {(256,  4.57) (512, 4.57) (1024, 4.92) (2048, 7.17)(4096, 11.67)};
        \addplot coordinates {(256, 4.43) (512, 4.43) (1024, 4.43) (2048, 4.43)(4096, 6.21)};
        \node[anchor=north west] at (rel axis cs:0.02,1) {\textbf{bsz=16}};
        
        \nextgroupplot[ymin=15,
        bar width=8pt]
        \addplot coordinates {(256, 15.5) (512, 15.5) (1024, 16.92) (2048, 0)(4096, 0)};
        \addplot coordinates {(256, 15.4) (512, 15.4) (1024, 15.4) (2048, 15.4)(4096, 0)};
        \node[anchor=north west] at (rel axis cs:0.02,1) {\textbf{bsz=64}};
        \node[blue, font=\fontsize{6}{7}\selectfont, align=center] at (rel axis cs:0.637,0.02) {oom};
        \node[blue, font=\fontsize{6}{7}\selectfont, align=center] at (rel axis cs:0.83,0.02) {oom};
        \node[red, font=\fontsize{6}{7}\selectfont, align=center] at (rel axis cs:0.94,0.02) {oom};
        
        \end{groupplot}
    \end{tikzpicture}
    \caption{Comparison of GPU memory consumption between Whisper and Whisper-MLA across different batch sizes and sequence lengths}
    \label{gpu memory}
\end{figure*}
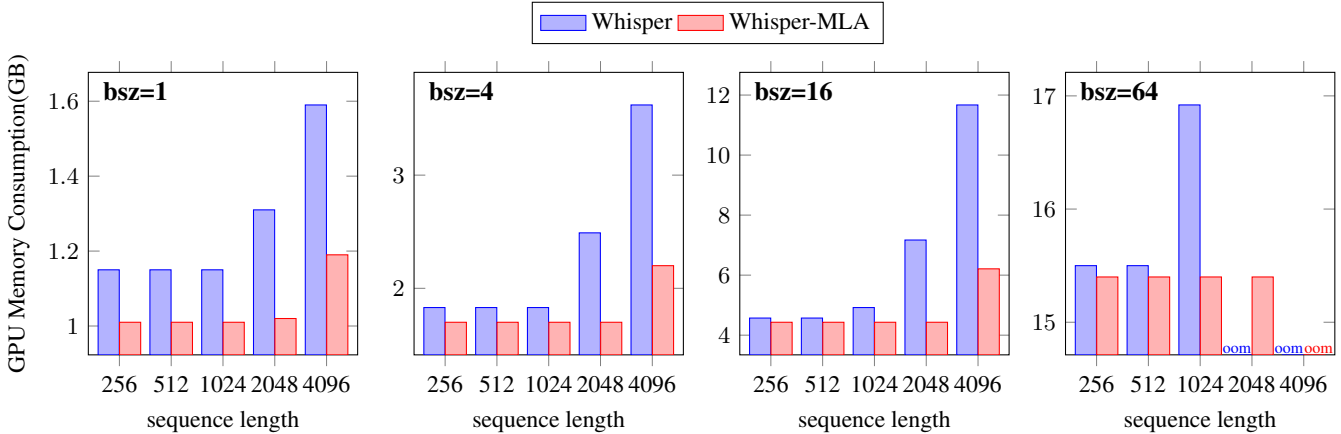

\section{Results and analysis}
\label{sec:result}

As shown in \autoref{tab:wer_comparison}, the Whisper-MLA models, generated via our proposed conversion and fine-tuning framework, successfully preserve the strong recognition capabilities of the original model. 
For a fair comparison, we also fine-tuned the standard Whisper model on the same dataset. Our results show that Whisper-MLA achieves a substantial KV cache reduction (up to 87.5\%) with minimal impact on ASR performance. 
Notably, the best-performing variant, Whisper-MLA (DSO) with uniform sampling, achieves a Word Error Rate (WER) that is highly competitive with the fine-tuned Whisper baseline, with an average WER across the four test sets that is only 0.17\% higher. This demonstrates an excellent trade-off between memory efficiency and accuracy.

Next, we compare the two architectural variants. Whisper-MLA (Full) shows reasonable ASR capability, confirming that MLA can be applied to encoder self-attention and cross-attention. However, its performance lags significantly behind Whisper-MLA (DSO). We attribute this to the DSO variant retaining more of the original pretrained parameters, especially in the encoder and cross-attention modules, which are critical for processing acoustic features. Consequently, we select the DSO architecture as our primary model. 

We further investigate the impact of different dimension preservation strategies. The results clearly show that preserving even a small subset of key dimensions (6.25\% of the total) substantially improves performance over full compression across all settings. For Whisper-MLA (Full), uniform and 2-norm strategies perform comparably. For Whisper-MLA (DSO), the uniform sampling strategy outperforms the 2-norm selection strategy. We infer this is because the Whisper decoder uses learnable positional embeddings, which lack the explicit frequency subspace structure of the encoder's sinusoidal embeddings. Thus, we select uniform sampling as the default dimension preservation strategy for Whisper-MLA.

Finally, we compare the GPU memory consumption of Whisper and Whisper-MLA under practical inference scenarios. As shown in \autoref{gpu memory}, Whisper-MLA consistently uses less memory. This gap becomes more pronounced as sequence length and batch size increase. 
For instance, at a batch size of 16 and a sequence length of 4096, Whisper-MLA consumes only about half of the memory of Whisper. 
More critically, when the batch size is 64 and the sequence length is 2048, Whisper's memory demand exceeds the 24GB GPU limit, resulting in out-of-memory (OOM), whereas Whisper-MLA operates comfortably at just 15.4 GB. These results underscore the superiority of Whisper-MLA for long-form speech recognition, especially with large batch sizes in resource-constrained environments.

\section{Conclusion}
\label{sec:conclude}

This paper presents Whisper-MLA, a novel architecture created through an efficient conversion for the Whisper model, successfully addressing its high memory consumption in ASR. Our approach features a unique adaptation of Multi-Head Latent Attention (MLA) for models with absolute positional embeddings. Through systematic placement experiments, we identify that applying MLA solely to the decoder's self-attention provides an optimal trade-off. These findings not only validate the effectiveness of our conversion methodology for encoder-decoder ASR architectures but also pave the way for deploying large-scale speech models on resource-constrained hardware, particularly for long-form audio applications.

\vfill\pagebreak

\section{Acknowledgement}
\label{sec:acknowledgement}

This work was supported in part by the National Key Research and Development Program of China under Grant 2023YFB2603902, in part by the Major Science and Technology Specific Project of Xining under Grant 2024-Z-7.

\bibliographystyle{IEEEbib}
\bibliography{refs}

@article{gulati2020conformer,
  title={Conformer: Convolution-augmented transformer for speech recognition},
  author={Gulati, Anmol and Qin, James and Chiu, Chung-Cheng and Parmar, Niki and Zhang, Yu and Yu, Jiahui and Han, Wei and Wang, Shibo and Zhang, Zhengdong and Wu, Yonghui and others},
  journal={arXiv preprint arXiv:2005.08100},
  year={2020}
}

@article{park2019specaugment,
  title={Specaugment: A simple data augmentation method for automatic speech recognition},
  author={Park, Daniel S and Chan, William and Zhang, Yu and Chiu, Chung-Cheng and Zoph, Barret and Cubuk, Ekin D and Le, Quoc V},
  journal={arXiv preprint arXiv:1904.08779},
  year={2019}
}

@inproceedings{dong2018speech,
  title={Speech-transformer: a no-recurrence sequence-to-sequence model for speech recognition},
  author={Dong, Linhao and Xu, Shuang and Xu, Bo},
  booktitle={2018 IEEE international conference on acoustics, speech and signal processing (ICASSP)},
  pages={5884--5888},
  year={2018},
  organization={IEEE}
}

@inproceedings{radford2023robust,
	title={Robust speech recognition via large-scale weak supervision},
	author={Radford, Alec and Kim, Jong Wook and Xu, Tao and Brockman, Greg and McLeavey, Christine and Sutskever, Ilya},
	booktitle={International conference on machine learning},
	pages={28492--28518},
	year={2023},
	organization={PMLR}
}

@article{10.1145/3530811,
author = {Tay, Yi and Dehghani, Mostafa and Bahri, Dara and Metzler, Donald},
title = {Efficient Transformers: A Survey},
year = {2022},
issue_date = {June 2023},
publisher = {Association for Computing Machinery},
address = {New York, NY, USA},
volume = {55},
number = {6},
issn = {0360-0300},
url = {https://doi.org/10.1145/3530811},
doi = {10.1145/3530811},
journal = {ACM Comput. Surv.},
month = dec,
articleno = {109},
numpages = {28}
}

@article{vaswani2017attention,
	title={Attention is all you need},
	author={Vaswani, Ashish and Shazeer, Noam and Parmar, Niki and Uszkoreit, Jakob and Jones, Llion and Gomez, Aidan N and Kaiser, {\L}ukasz and Polosukhin, Illia},
	journal={Advances in neural information processing systems},
	volume={30},
	year={2017}
}

@inproceedings{kwon2023efficient,
  title={Efficient memory management for large language model serving with pagedattention},
  author={Kwon, Woosuk and Li, Zhuohan and Zhuang, Siyuan and Sheng, Ying and Zheng, Lianmin and Yu, Cody Hao and Gonzalez, Joseph and Zhang, Hao and Stoica, Ion},
  booktitle={Proceedings of the 29th symposium on operating systems principles},
  pages={611--626},
  year={2023}
}

@article{liu2024deepseek,
	title={Deepseek-v2: A strong, economical, and efficient mixture-of-experts language model},
	author={Liu, Aixin and Feng, Bei and Wang, Bin and Wang, Bingxuan and Liu, Bo and Zhao, Chenggang and Dengr, Chengqi and Ruan, Chong and Dai, Damai and Guo, Daya and others},
	journal={arXiv preprint arXiv:2405.04434},
	year={2024}
}

@article{ji2025towards,
	title={Towards Economical Inference: Enabling DeepSeek's Multi-Head Latent Attention in Any Transformer-based LLMs},
	author={Ji, Tao and Guo, Bin and Wu, Yuanbin and Guo, Qipeng and Shen, Lixing and Chen, Zhan and Qiu, Xipeng and Zhang, Qi and Gui, Tao},
	journal={arXiv preprint arXiv:2502.14837},
	year={2025}
}

@inproceedings{zhang2020transformer,
  title={Transformer transducer: A streamable speech recognition model with transformer encoders and rnn-t loss},
  author={Zhang, Qian and Lu, Han and Sak, Hasim and Tripathi, Anshuman and McDermott, Erik and Koo, Stephen and Kumar, Shankar},
  booktitle={ICASSP 2020-2020 IEEE International Conference on Acoustics, Speech and Signal Processing (ICASSP)},
  pages={7829--7833},
  year={2020},
  organization={IEEE}
}

@article{touvron2023llama,
  title={Llama 2: Open foundation and fine-tuned chat models},
  author={Touvron, Hugo and Martin, Louis and Stone, Kevin and Albert, Peter and Almahairi, Amjad and Babaei, Yasmine and Bashlykov, Nikolay and Batra, Soumya and Bhargava, Prajjwal and Bhosale, Shruti and others},
  journal={arXiv preprint arXiv:2307.09288},
  year={2023}
}

@article{bai2023qwen,
  title={Qwen technical report},
  author={Bai, Jinze and Bai, Shuai and Chu, Yunfei and Cui, Zeyu and Dang, Kai and Deng, Xiaodong and Fan, Yang and Ge, Wenbin and Han, Yu and Huang, Fei and others},
  journal={arXiv preprint arXiv:2309.16609},
  year={2023}
}

@article{su2024roformer,
	title={Roformer: Enhanced transformer with rotary position embedding},
	author={Su, Jianlin and Ahmed, Murtadha and Lu, Yu and Pan, Shengfeng and Bo, Wen and Liu, Yunfeng},
	journal={Neurocomputing},
	volume={568},
	pages={127063},
	year={2024},
	publisher={Elsevier}
}

@article{barbero2410round,
	title={Round and round we go! what makes rotary positional encodings useful?},
	author={Barbero, Federico and Vitvitskyi, Alex and Perivolaropoulos, Christos and Pascanu, Razvan and Veli{\v{c}}kovi{\'c}, Petar},
	journal={URL https://arxiv. org/abs/2410.06205},
        year={2025}
}

@article{tang2023salmonn,
  title={Salmonn: Towards generic hearing abilities for large language models},
  author={Tang, Changli and Yu, Wenyi and Sun, Guangzhi and Chen, Xianzhao and Tan, Tian and Li, Wei and Lu, Lu and Ma, Zejun and Zhang, Chao},
  journal={arXiv preprint arXiv:2310.13289},
  year={2023}
}

@article{xie2024mini,
  title={Mini-omni: Language models can hear, talk while thinking in streaming},
  author={Xie, Zhifei and Wu, Changqiao},
  journal={arXiv preprint arXiv:2408.16725},
  year={2024}
}

@article{fang2024llama,
  title={Llama-omni: Seamless speech interaction with large language models},
  author={Fang, Qingkai and Guo, Shoutao and Zhou, Yan and Ma, Zhengrui and Zhang, Shaolei and Feng, Yang},
  journal={arXiv preprint arXiv:2409.06666},
  year={2024}
}

@inproceedings{panayotov2015librispeech,
	title={Librispeech: an asr corpus based on public domain audio books},
	author={Panayotov, Vassil and Chen, Guoguo and Povey, Daniel and Khudanpur, Sanjeev},
	booktitle={2015 IEEE international conference on acoustics, speech and signal processing (ICASSP)},
	pages={5206--5210},
	year={2015},
	organization={IEEE}
}

\end{document}